\documentclass{article}

\usepackage{spconf,amsmath,graphicx,cite}

\usepackage{subfig}
\usepackage{bm}
\usepackage{float}
\usepackage{xcolor}
\usepackage{balance}

\usepackage{amsfonts}
\usepackage{diagbox}
\usepackage{color}
\usepackage{url}

\title{Cross-Attention is all you need: Real-Time Streaming Transformers for Personalised Speech Enhancement}
\name{Shucong Zhang, Malcolm Chadwick, Alberto Gil C. P. Ramos, Sourav Bhattacharya}
\address{Samsung AI Centre, Cambridge, UK}
\begin{document}
%
\maketitle

\begin{abstract}
Personalised speech enhancement (PSE), which extracts only the speech of a target user and removes everything else from a recorded audio clip, can potentially improve users' experiences of audio AI modules deployed in the wild. To support a large variety of downstream audio tasks, such as real-time ASR and audio-call enhancement, a PSE solution should operate in a streaming mode, i.e., input audio cleaning should happen in real-time with a small latency and real-time factor. Personalisation is typically achieved by extracting a target speaker's voice profile from an enrolment audio, in the form of a static embedding vector, and then using it to condition the output of a PSE model. However, a fixed target speaker embedding may not be optimal under all conditions. In this work, we present a streaming Transformer-based PSE model and propose a novel cross-attention approach that gives adaptive target speaker representations. We present extensive experiments and show that our proposed cross-attention approach outperforms competitive baselines consistently, even when our model is only approximately half the size. 

\end{abstract}
\begin{keywords}
Speech separation, speech enhancement, personalisation, deep learning, cross attention
\end{keywords}
\vspace{-1mm}
\section{Introduction}
\label{sec:intro}
\vspace{-2mm}
Audio source separation is an important problem within the audio research community and applications  of deep neural networks have lately shown great performance improvements~\cite{wang2018voicefilter, luo2019conv, wang2020voicefilterlite, chen2020dual}. 
In this work, we focus on a sub-problem of source separation, personalised speech enhancement (PSE), where the objective is to extract the speech of a target user from a noisy, single channel, digital audio recording. All other audio components, such as environmental noise and other voices, are removed.
Personalisation is performed by extracting speaker embeddings, which capture the voice profile of a user, from previously recorded short-enrolment audio of that user, e.g., ``Hi Bixby''. 
These embeddings are then used during inference to condition the output of a PSE model.

\noindent
PSE has great potential in improving user experiences of audio AI modules deployed in the wild, e.g., improving the call-quality during telephony, reducing recognition error rates of a downstream ASR when applied in a noisy environment, and enhancing audio interactability of home-robots. 
We further focus on online PSE approaches -- in particular streaming models -- which can be used as a pre-processing module for a number of downstream AI modules. 


\noindent
A common approach in PSE is to use a separate embedding network to extract a speaker embedding, often only a single vector, for the target speaker~\cite{wang2018voicefilter}. This provides a constant pre-computed embedding for each speaker, which improves run-time efficiency. During inference, the extracted speaker embedding is concatenated with the intermediate features of the PSE network for conditioning~\cite{wang2018voicefilter, wang2020voicefilterlite}.
Other advanced approaches use the constant speaker embedding vector through feature modulation \cite{o2021conformer, Gfeller2021oneshot} or by learning activation functions~\cite{ramos2022conditioning} for the purpose of personalisation.
\vspace{1mm}

%

\noindent
Despite the success of previous PSE models, we argue that since the acoustic characteristics, e.g, speech, pitch and emotion, of the target speaker may vary across time and/or across different utterances, a fixed constant vector may not be an optimal voice profile. Thus, we explore a learning-based cross-attention approach, which uses multiple hidden states from a pre-trained speaker embedding model to produce an adaptive embedding representation for the current noisy audio frame. 
To the best of our knowledge, cross-attention is used for the first time in this work to generate adaptive target speaker embeddings. For the PSE backbone, we develop a fully attention-based streaming Transformer model. Through extensive experiments, we show our proposed cross-attention models consistently outperform baseline models,
both in PSE and downstream ASR tasks, even when only half the size.

\vspace{-3mm}

\section{Related Work}
\vspace{-2mm}
\label{sec:related}


A critical task in PSE is to make the model aware of the acoustic characteristics of a target speaker. 
To achieve this, previous works extract a single fixed speaker embedding vector, such as an i-vector \cite{dehak2010front} or d-vector \cite{wan2018generalized}, from enrolment audio. 
This static vector can be used in multiple ways: \cite{wang2018voicefilter, wang2020voicefilterlite, Xu2019tdsen, li2020pseattn, ji2020joint, Eskimez2022newpse, Giri2021percepNet, Manthan2022e3} concatenate it with the corrupted input hidden representations, while \cite{o2021conformer, Gfeller2021oneshot} use FiLM \cite{Ethan2018film} to integrate it with the model. In contrast, we propose to use cross-attention to learn adaptive, rather than static, representations of the target speaker.

\vspace{1mm}
\noindent
While there exist PSE solutions with attention components, most of them do not leverage attention mechanisms to generate adaptive voice representations of the target speaker. \cite{wang2019speech, rikhye2022closing} use attention mechanisms to select a speaker embedding from an inventory of fixed embeddings of different speakers. \cite{ han2021attention,hu2021avatr} use attention mechanisms to inject speaker's voice profile into PSE models, but this was achieved by repeating the target speaker's single constant embedding vector to match the length of the input.
\vspace{1mm}

\noindent
\cite{xiao2019single} applies one attention layer to  the enrolment utterances of the target and interference speaker, but applying attention only to the target speaker is not explored. Also in \cite{xiao2019single}, the outputs of the attention layer are still concatenated with the hidden representations of the  inputs and fed into a Conv-TasNet \cite{luo2019conv}. In this work, we are the first to propose a novel way to use cross-attention for adaptive representations of the target speaker in a solely attention-based PSE model. 

\vspace{-4mm}
\section{Method}
\label{sec:Method}
\vspace{-2mm}
Given a corrupted audio sequence $\bm{X}'$ and the voice profile of a target speaker,
the task is to extract the target speaker's speech $\bm{X}$
accurately. 
Here, we consider PSE as a spectral-subtraction task~\cite{vaseghi2008advanced} and thus both $\bm{X}'$ and $\bm{X}$ are spectrograms. 
We propose a streaming Transformer-based encoder-decoder model and Fig.~\ref{fig:model} presents an architectural overview of the proposed model and the baseline model.

\vspace{-3mm}

\vspace{-1mm}
\subsection{Self-attention Encoder}
\begin{figure}[]
\centering
\subfloat[Proposed Model]{\includegraphics[width=0.42\columnwidth]{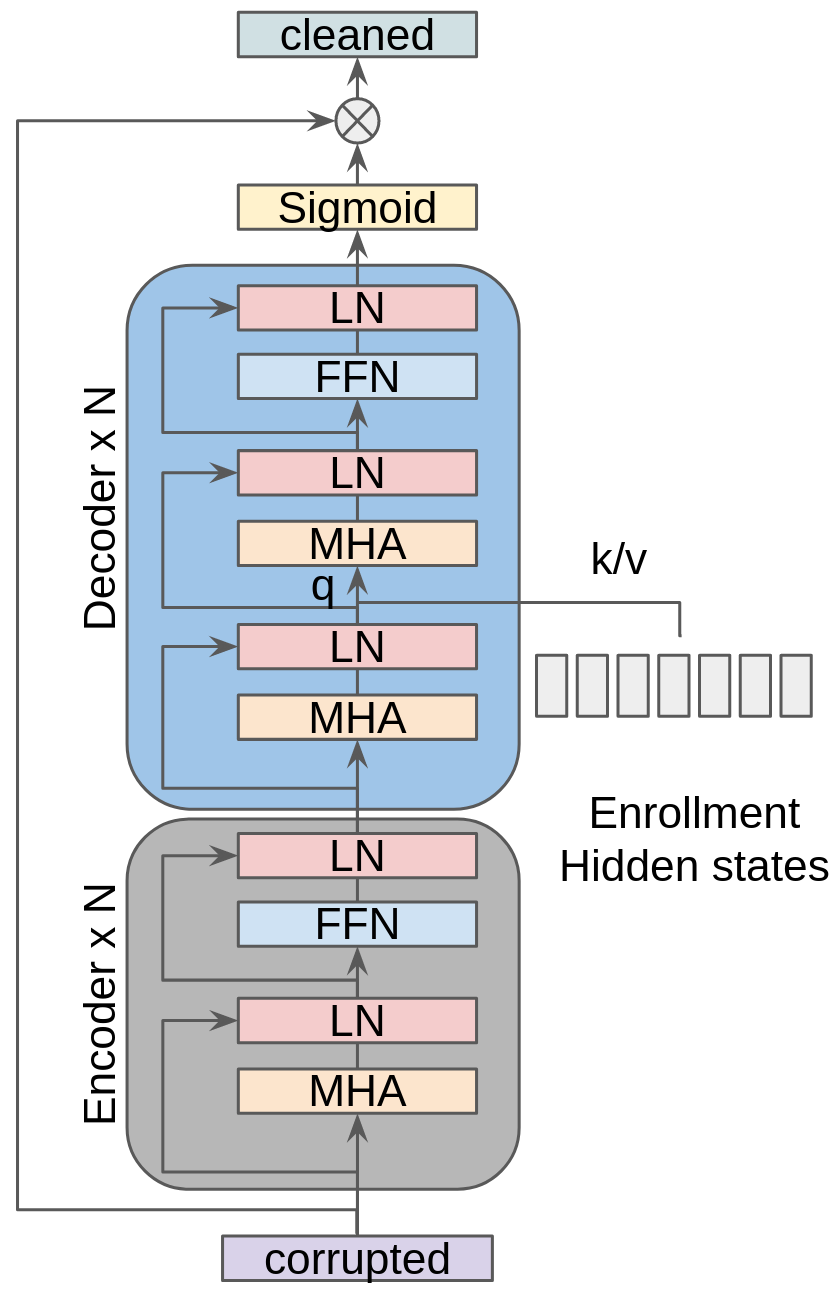}}
\qquad
\subfloat[Baseline Model]{\includegraphics[width=0.4\columnwidth]{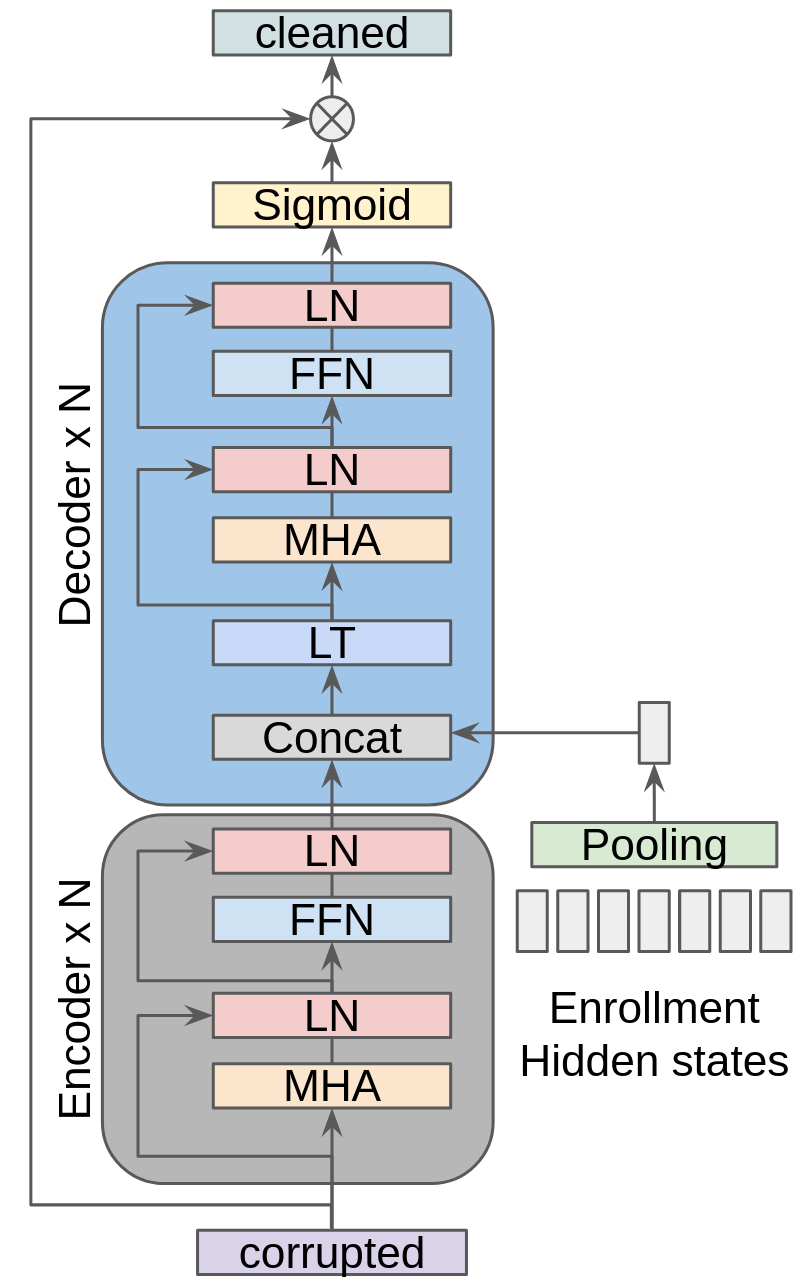}}
\caption{Architectural overview: (a) cross-attention decoder, and (b) concatenation-based decoder (baseline). \textbf{MHA}, \textbf{LN}, $\textbf{FFN}$, \textbf{Concat}, and \textbf{LT} denote multi-headed attention \cite{vaswani2017attention}, layer normalisation \cite{ba2016layer}, feed-forward network, concatenation along the channel axis and linear transformation respectively.
\textbf{Pooling} represents either average or last pooling.
}
\vspace{-4mm}
\label{fig:model}
\end{figure}

\vspace{-1mm}
We use a self-attention encoder to transform a corrupted input $\bm{X}'=(\bm{x}'_1, \bm{x}'_2,...,\bm{x}'_t)$ to a sequence of hidden states $\bm{Z}=(\bm{z}_1, \bm{z}_2,...,\bm{z}_t)$, where $\bm{X}'\in \mathbb{R}^{t \times d_{fft}}$ and $\bm{Z} \in \mathbb{R}^{t \times d_{model}}$ for both architectures (see Fig.~\ref{fig:model}). To achieve real-time streaming, we apply a relative positional encoding (RPE)~\cite{shaw18relative, dai2019transformer} and a mask for the MHA component 
such that at each time step, it only looks back $k$-history frames. 

\vspace{-2mm}
\subsection{Concatenation-based Decoder: Baseline}
\vspace{-1mm}
We use a pre-trained d-vector model~\cite{wan2018generalized} to transform the enrolment audio $\bm{A}$ of the target speaker to a sequence of hidden states $\bm{H}$. Then, we apply an average pooling or a last pooling along the time axis of $\bm{H}$ to get a hidden representation $\bm{h}$ as the target speaker's voice profile. We also consider the scenario where multiple enrolment utterances $\bm{A}^{1-q}$ are available. In this setting, we apply the pre-trained d-vector model to each enrolment utterance to get $\bm{H}^{1-q}$, and then use either an average pooling or a last pooling to get $q$ hidden representations $\bm{h}^{1-q}$. The average of $\bm{h}^{1-q}$ is used as the target speaker's voice profile. We denote the constant target speaker hidden representation from either the single enrolment utterance or the multiple enrolment utterances as $\bm{h}_F$. As the main baseline approach (see Fig.~\ref{fig:model}b), we consider feature concatenation, where $\bm{h}_F$ is concatenated with each of the hidden representations of the corrupted audio for solving the PSE task. 
Formally, suppose there are $n$ layers in the decoder, and the input for each layer is $\bm{Y}$, where $\bm{Y}$ is the encoder hidden states $\bm{Z}$ or the output of the previous decoder layer, then the operations within each decoder layer can be described as:
\vspace{-1mm}
\begin{align}
    \bm{Y}' &= \mathrm{LT}(\mathrm{Concat}(\bm{Y},\,\bm{h}_F))\\
    \bm{Y}'' &= \mathrm{LN}(\bm{Y}' + \mathrm{MHA}(Q=\bm{Y}', K=\bm{Y}', V=\bm{Y}'))\\
    \bm{Y}''' &= \mathrm{LN}(\bm{Y}'' + \mathrm{FFN}(\bm{Y}''))
\end{align}
where $\mathrm{Concat}(\cdot)$, $\mathrm{LT}(\cdot)$, $\mathrm{FFN}(\cdot)$, $\mathrm{LN}(\cdot)$ and $\mathrm{MHA}(\cdot)$ are as defined in the caption of Fig.~\ref{fig:model}.
In particular, $\mathrm{LT}$ projects $(\bm{y}_i,\bm{h}_F)$ from $\mathbb{R}^{d_{model}+d_{embed}}$ to $\mathbb{R}^{d_{model}}$. 
We also apply RPE and masking for each decoder MHA. The output of the decoder is transformed back to $\mathbb{R}^{t \times d_{fft}}$ by applying a linear transformation and then fed into a Sigmoid function to generate a mask, which is then applied to $\bm{X}'$ to generate the extracted output $\bm{X}$.

\vspace{-3mm}
\subsection{Cross-attention Decoder: Proposed Solution}
\vspace{-1mm}
Although highly successful, we argue that the constant fixed speaker embedding may not always best match the target speaker's voice features, which may vary across different time steps or across different utterances during deployment. Therefore, motivated by attention-based neural machine translation models \cite{bahdanau2014neural, vaswani2017attention}, we use cross-attention to dynamically generate the target speaker's voice profile, which may better fit the target speaker's voice characteristic at the current input time step. As shown in Fig.~\ref{fig:model}(a), we view the hidden representation of the current input frame as the Query. We view $\bm{H}$ from the single enrolment audio or $\bm{h}^{1-q}$ from the multiple enrolment utterances as the Key/Value for computing the cross-attention. Therefore, based on the current input frame, the cross-attention learns to do either a ``soft selection'' of suitable enrolment hidden states along the time axis of the single enrolment audio, or a ``soft selection'' of the most suitable enrolment audio from multiple enrolment utterances.
We denote $\bm{H}$/$\bm{h}^{1-q}$ from the single/multiple enrolment utterance(s) as $\bm{H}_F$. We use an additional linear transformation to map the channel dimension of $\bm{H}_F$ to the channel dimension of $\bm{Y}$. 
Denoting the output of the additional linear transformation as $\bm{H}'$, our proposed cross-attention decoder can be described as:
\vspace{-1mm}
\begin{align}
    \bm{Y}' &= \mathrm{LN}(\bm{Y} + \mathrm{MHA}(Q=\bm{Y}, K=\bm{Y}, V=\bm{Y})) \\
    \bm{Y}'' &= \mathrm{LN}(\bm{Y}' + \mathrm{MHA}(Q=\bm{Y}', K=\bm{H'}, V=\bm{H'})) \\
    \bm{Y}''' &= \mathrm{LN}(\bm{Y}'' + \mathrm{FFN}(\bm{Y}'')).
\end{align} We apply masking and RPE to the first MHA component to enable streaming.
However, for the multi-headed cross-attention between $\bm{Y}'$ and $\bm{H}'$, since $\bm{H}'$ is the hidden states of the enrolment audio(s), which are usually pre-computed and available locally, we do not need to apply any masking to support streaming.

\vspace{-3mm}
\subsection{Time and Space Complexity}
\vspace{-1mm}
Given $\bm{A} \in \mathbb{R}^{d_i \times d_j}$ and $\bm{B} \in \mathbb{R}^{d_j \times d_k}$, we assume the time complexity for computing $\bm{A} \bm{B}$ is $O(d_i d_j d_k)$. 
By pre-computing and caching the linear transformations of the enrolment utterance(s) hidden states, $\mathrm{LT}(\bm{y}_i,\bm{h}_F)$ will cost $O(d_{model}^2)$, and  $\mathrm{MHA}(\bm{Y}', \bm{H'}, \bm{H'})$ will cost $O(d_{model}^2)$ for the linear transformation of $\bm{Y}'$ and $O(p \cdot d_{model})$ for the dot-products, 
where $p$ is the number of frames in the single enrolment utterance case, and the number of utterances in the multiple utterance case. 
Especially, in the multiple utterance case, typically $p \ll d_{model}$, and thus, compared with the baseline, our approach has the same time complexity of $O(d_{model}^2)$, with a negligible increase in space requirements due to caching. 
However, this minor difference is greatly offset by the improved modelling capabilities (see Table~\ref{tab:results}), where we find that with only half the parameters our proposed models can give similar or superior results than the baselines. 
\vspace{1mm}



\begin{table*}[h]
\caption{The mean and standard deviation of both the signal-to-distortion-ratio (SDR, higher is better) and word error rate (WER, lower is better), across models obtained by different training. The WER is calculated using an open-source STT engine \cite{silero-models}. \#Layers/\#Param denote the total number of layers/parameters
of the model. We train each model 3 to 9 times to make sure the improvement is statistically significant. 
$\dagger$ indicates $\operatorname{p-value} < 0.025$ and  $\ddagger$ means $\operatorname{p-value}< 0.01$ by one-tailed t-tests for cross-attention v.s. mean pooling in SDR. For the Babble WER improvements the $\operatorname{p-value}$ varies from 0.07 to 0.003.} 
\vspace{-2mm}
\label{tab:results}
\centering
\begin{tabular}{l|c|c|c|c|c|r}
\hline
Model    & Dev SDR   & Eval SDR & Ambient WER & Babble WER & \#Layers & \#Param.  \\ \hline
Ground Truth & N/A & N/A & 8.34  & 8.03 & N/A  & N/A                   \\ 
No PSE & N/A & N/A & 18.32  & 72.32 & N/A  & N/A
\\
VoiceFilter (bidirectional) & 16.71 & 16.09 & 14.31 & 22.72 & N/A  & 7.8M \\ 
Cross-attention (bidirectional) & 18.70 & 18.16 & 15.43 & 21.68 & 6  & 6.1M \\
{Cross-attention (bidirectional)} & \textbf{18.96} & \textbf{18.41} & \textbf{13.72}  & \textbf{17.62} & 12  & 12.0M \\ \hline
 \multicolumn{7}{c}{One enrolment utterance}\\ \hline

Last Pooling Concat & 15.31 $\pm$ 0.05 & 14.94 $\pm$ 0.07 & 15.85 $\pm$ 0.13 & 28.90 $\pm$ 1.10 & 6           & 5.8M                   \\
Mean Pooling Concat  & 15.58 $\pm$ 0.17 & 15.16 $\pm$ 0.09 & 15.68 $\pm$ 0.12 & 26.56 $\pm$ 0.51 & 6           & 5.8M                   \\
Cross-attention $\dagger$ & \textbf{15.71} $\pm$ 0.08 & \textbf{15.29} $\pm$ 0.06 & \textbf{15.66} $\pm$ 0.15
 & \textbf{26.03} $\pm$ 0.51 & 6 & 6.1M                \\ \hline
Last Pooling  Concat & 15.64 $\pm$ 0.03   & 15.23 $\pm$ 0.06 & 15.41 $\pm$ 0.11 & 25.87 $\pm$ 0.88 & 12   & 11.3M               \\
Mean Pooling Concat & 15.84 $\pm$ 0.09  & 15.44 $\pm$ 0.05 & 15.14 $\pm$ 0.24 & 23.73 $\pm$ 0.63 & 12    & 11.3M                   \\
Cross-attention $\ddagger$ & \textbf{16.07} $\pm$ 0.07 & \textbf{15.62} $\pm$ 0.09 & \textbf{14.98} $\pm$ 0.07 & \textbf{22.24} $\pm$ 0.67 & 12      & 12.0M          \\
\hline 
\multicolumn{7}{c}{Five enrolment utterances}\\ \hline
Mean Last Pooling Concat & 15.62 $\pm$ 0.14 & 15.22 $\pm$ 0.15  & 15.99 $\pm$ 0.22 & 26.48 $\pm$ 1.18 & 6 & 5.8M                   \\
Cross-attention $\ddagger$ & \textbf{15.93} $\pm$ 0.17 & \textbf{15.48} $\pm$ 0.22 & \textbf{15.82} $\pm$ 0.27 & \textbf{25.11} $\pm$ 1.44 & 6 & 6.1M                \\ \hline
Mean Last Pooling Concat & 15.91 $\pm$ 0.08  & 15.54 $\pm$ 0.08 & 15.65 $\pm$ 0.10 & 23.65 $\pm$ 0.80 & 12 & 11.3M             \\
Cross-attention $\ddagger$ & \textbf{16.23} $\pm$ 0.10 & \textbf{15.84} $\pm$ 0.05 & \textbf{15.51} $\pm$ 0.08 & \textbf{22.65} $\pm$ 0.08 & 12  & 12.0M          \\
\hline
\end{tabular}
\end{table*}
\vspace{-8mm}
\section{Experiments}
\vspace{-2mm}
\noindent

\noindent
\textbf{Dataset.} For ground-truth clean speech we consider the clean splits of LibriSpeech \cite{panayotov2015librispeech}.
For each clean audio, we pair it with either an ambient noise (45\%) from FreeSoundDataset~\cite{freesound} or a babble noise (45\%) from a different speaker in LibriSpeech. The remaining 10\% are left as clean samples, with a small amount of white noise added. The target speech and noise are then added to create the corrupted data, with the noise amplitude adjusted to create varying SNR values (sampled uniformly between ${-3}$ and $10$ inclusive). Each corrupted audio is then matched with 1-5 enrolment utterances, which are random audio clips from the same speaker in the LibriSpeech dataset. For each utterance, a random 3 second chunk (after removing silence) is taken from the longer clip to act as a short ``Hi Bixby'' equivalent. The corrupted clips are then segmented into 3 second chunks, to allow for efficient batching and training. For calculating WER, the non chunked audio is used instead to avoid transcription issues.
We plan to release our validation and test datasets.
\vspace{1mm}

\noindent
\textbf{Experimental Setup.} To investigate the performance of our approach at different scales, we built two different sizes of Transformer PSE models. For the base model, both the encoder and decoder have 3 layers each. For the large model, they have 6 layers each. All MHA components have 8 heads and each head has dimension 32. The FFN has one hidden layer of dimension 1024. We apply dropout \cite{srivastava2014dropout} with a probability 0.1 to the output of each sub-component in each layer. For streaming, we restrict the look back step to 100 frames.
We train the base/larger models for 100/200 epochs with batch size 320/640. 
Each epoch contains 350 hours of training data. We follow the optimizer in \cite{vaswani2017attention} with 16k warm-up steps. The training objective is to reduce the power-law compressed spectral distance \cite{braun2021consolidated} between the cleaned spectrograms produced by the models as in Fig.\ref{fig:model} and the ground-truth.
We also implement and train a bidirectional VoiceFilter model as a baseline following the model architecture in \cite{wang2018voicefilter}, and we also train bidirectional cross-attention models. 
We follow \cite{wan2018generalized} to build a d-vector model and use it to generate the hidden states of the enrolment audio(s). Models are implemented via Tensorflow \cite{abadi2016tensorflow} and Horovod \cite{sergeev2018horovod}. 
\vspace{1mm}

\noindent
\textbf{Experimental Results.}
To put our results in Table~\ref{tab:results} into context, we consider various baselines and competing methods.
 Firstly, we consider the case of no corruption, where the ground-truth audio has not been augmented, as a quality upper bound. 
Secondly, we consider the case of no enhancement, where the corrupted audio is not denoised by the PSE model, as a quality lower bound. Thirdly, although we are primarily interested in streaming scenarios, we provide bidirectional non-streaming PSE models which can be seen as a soft upper bound.
Here, our non-streaming cross-attention models 
outperform the non-streaming VoiceFilter baseline by a large margin, even with fewer parameters.
Our main investigation compares the proposed cross-attention approach against various (last or mean) pooling concatenation baselines, where the systems have access to either one or five enrolment utterances. Impressively in both deployment scenarios, our large streaming cross-attention models give similar or lower WER for babble noise compared to the non-streaming VoiceFilter, which has access to future information.
\vspace{1mm}

\noindent
For the enrolment case with a single utterance, the mean pooling approach yields better results than the last pooling approach, and our proposed cross-attention method surpasses both. This indicates the target speaker's acoustic features indeed tend to vary through time, and the adaptive target speaker's voice embedding generated by our method is more helpful than the constant fixed embedding.
To exclude the possibility that improvements solely come from the more complicated cross-attention architecture, we repeat the static speaker embedding, as in \cite{hu2021avatr}, to execute the cross-attention, and we obtain similar results to the baselines.
\vspace{1mm}


\noindent
For the enrolment scenario with five utterances, we apply either mean pooling then concatenation or cross-attention to the individual utterance embeddings $\bm{h}^{1-5}$.  
We find that obtaining the individual utterance embedding $\bm{h}^{i}$ from $\bm{H}^i$ through last pooling or average pooling makes little difference to the performance of the PSE models. 
This implies that the information across different utterances is significantly more diverse than the information across different time steps within the same utterance.
Our cross-attention approach consistently gives the best results, which shows that the cross-attention is able to ``soft select'' the most suitable embeddings across different utterances, to best match the current streamed corrupted input frame. Although the models trained with five utterances yield higher SDR than the models trained with one, they do not have lower WERs. The mismatch between SDR and WER is a common problem and addressing this is beyond the scope of this paper. We also note that during real deployment, multiple utterances for enrolment may not always be available. However, we can use data argumentation to generate multiple utterances from a single one, and we leave this as a future work. 
\vspace{0.5mm}

\noindent
In both of these enrolment scenarios, the WER reduction from cross-attention in the Babble noise condition is much larger than the Ambient noise condition. This is expected since the background Ambient noise differs significantly from the target speaker's voice and thus, a rough target speaker representation could be sufficient for PSE. Furthermore, most ASR systems are trained with data augmentation to increase robustness to ambient noise.
In the Babble noise condition however, the PSE model needs precise information about the target speaker's voice characteristic to separate the target speaker from the interference speaker and thus, the cross-attention approach shows a larger reduction in WER. Finally, it is worth noting that for both enrolment scenarios, with approximately half the model size, our base cross-attention models give similar or better results compared to the large concatenation baselines. 



\vspace{-6mm}
\section{Conclusion}
\label{sec:conclusion}
\vspace{-2mm}
We proposed a streaming and fully Transformer-based PSE model that builds on a novel cross-attention approach to dynamically generate suitable target speaker embeddings. This adaptability boosts the enhancement of the incoming corrupted audio from a single or limited number of enrolment utterances such as ``Hi Bixby''. 
This is in stark contrast with existing approaches that extract a fixed speaker embedding from the enrolment data, effectively averaging out any variation in the enrolment that may be useful at inference.
Through a large number of 
experiments, we show that our proposed method consistently surpasses the static target speaker embedding approach in both PSE and downstream ASR tasks.
Furthermore, in the non-streaming setting, which is not the main focus of the paper, we find that our proposed approach yields significantly better results than the VoiceFilter (offline) model. 
Finally, although we focus on fully attention based models in this paper, it is straightforward to apply our proposed cross-attention component to inject adaptive target speaker embeddings to PSE models with other architectures, which we leave as a future work. Adapting our model to the raw-wave inputs is also an interesting direction.





%
%
%

\bibliographystyle{IEEEtran}
\ninept
\balance
\bibliography{strings,refs}

\end{document}